\documentstyle[preprint,aps]{revtex}

\begin{document} \draft \title{Classical spin contribution to the Tunnel 
Effect} \author{ Mart\'{\i}n\ Rivas} \address{Dpto. de F\'{\i}sica 
Te\' orica, Universidad del Pa\'{\i}s Vasco-Euskal Herriko 
Unibertsitatea,\\ Apdo.~644, 48080 Bilbao, Spain \\ E-mail: 
wtpripem@lg.ehu.es} 

\date{\today}

\maketitle

 \begin{abstract} Since the spin of real particles is of order of 
$\hbar$, it is difficult to distinguish in a quantum mechanical 
experiment involving spinning particles what part of the outcome is 
related to the spin contribution and what part is a pure quantum 
mechanical effect. We analyze in detail a classical model of a 
nonrelativistic spinning particle under the action of a potential 
barrier and compute numerically the crossing for different potentials. 
In this way it is shown that because of the spin structure there is a 
nonvanishing contribution to crossing for energies above a certain 
minimum value, even below the top of the potential barrier. Results 
are compared with the quantum tunnel effect.
 \end{abstract}

\vspace{2cm}
\newpage

\section{Introduction}
\label{sec:intro}

The aim of this work is to describe in classical terms the classical 
spin contribution to the so called tunnel effect by assuming a 
particular spin structure of a classical spinning particle. 

In section~\ref{sec:limit} we make some general considerations about 
the classical limit of quantum mechanics, and also about the 
definition of elementary particle in quantum mechanics and the 
different treatment deserved to spin in classical as well as in 
quantum mechanics. These considerations lead us to propose a general 
definition of elementary particle at the classical level that 
resembles the one used in quantum mechanics. 

A Lagrangian approach of a nonrelativistic spinning particle model to 
be used later, is described in detail in Sec.~\ref{sec:particle} where 
the main features are that the particle spin arises because the 
separation between its center of mass and center of charge. The center 
of mass satisfies Newton's equations where the external force is 
defined on the point charge position while this center of charge moves 
around the center of mass with an isotropic harmonic motion giving 
rise to the magnetic properties of the particle. The canonical 
approach of this model is also presented in Sec.~\ref{sec:canonical}. 

We shall consider the solutions of a one-dimensional model of this 
classical spinning particle interacting with a potential barrier in 
Section~\ref{sec:tunel}. Dynamical equations are nonlinear and a 
general solution is dificult to obtain in closed form even when the 
external potential is a linear function of the charge position. We 
solve the problem by numerical integration showing the penetration 
effect for different initial conditions and different slopes of the 
potential barrier. The crossing is achieved because of the separation 
between the center of mass and center of charge in such a way that the 
average repulsive force acting on the spinning particle is smaller 
than in the point particle case. Results are compared with the quantum 
tunnel effect for a point particle under the same one-dimensional 
potential.

\section{Classical Limit of Quantum Mechanics}
\label{sec:limit}

One usually reads in textbooks on quantum mechanics that {\sl 
Classical Mechanics} (CM) {\sl is the limit} $\hbar\to0$ {\sl of 
Quantum Mechanics} (QM)\cite{ref1}. Of course there are exceptions. 
For instance in Feynman's Lectures on Physics\cite{Feynman} we read: 
`{\sl In the classical limit, the quantum mechanics will agree with 
Newtonian Mechanics}'. Both expressions are equivalent if by CM we 
understand the classical mechanics of spinless (or point) particles, 
i.e., Newtonian Mechanics. We think this is the meaning the mentioned 
authors try to express. But the first statement as it stands, might 
lead to wrong interpretations if considered literally, as some of the 
quoted references might suggest. In order to clarify this idea of the 
classical limit of QM, let us consider the following simplified 
diagram of Fig.~\ref{fig:fig1}.

The set $A$ represents the whole body of knowledge of CM which 
includes two subsets, the subset $(s=0)$ or domain of spinless 
particles or Newtonian Mechanics and the subset $(s\neq0)$ of spinning 
particles. If we restrict CM to satisfy the additional requirement of 
the Uncertainty Principle we enter into the more restricted body of 
knowledge of QM, represented by the smaller set $B$ in which we also 
have the two subdomains of spinless and spinning particles. If in this 
more constrained domain we perform now the additional restriction of 
taking the limit $\hbar\to0$ it is out of logic that after these two 
restrictions we shall reach the wider and less restrictive domain of 
the whole CM. 

In QM we have that the measurement of any two observables $C$ and $D$ 
is not in general compatible, and the uncertainty in their 
simultaneous measurement is related to its commutator 
$[C,D]\simeq\hbar$ wich is of order of Planck's constant $\hbar$. But 
we also have in QM that the particle states satisfy eigenvalue 
equations for the spin of the form $S^2|\psi>=s(s+1)\hbar^2|\psi>$, 
where the right hand side is also a function of $\hbar$. It turns out 
that when performing the limit $\hbar\to0$ we get $[C,D]=0$ and also 
$S^2|\psi>=0$, i.e., the physics of compatible observables of spinless 
systems. 

With this analysis we see that the limit $\hbar\to0$ of QM is in fact 
Newtonian Mechanics and not the whole domain of CM. Perhaps one of the 
reasons for the identification of CM with Newtonian Mechanics in the 
mentioned references lies in the fact that we are used to work in QM 
with spinning systems since the early days of the quantum theory while 
the CM of spinning particles is still waiting for a complete 
development and improvement at least equivalent to the one we have 
achieved in the quantum domain. Even we can remember here that for 
many years, spin has been considered by physicists a strict quantum 
mechanical and relativistic property of the electron, as was pointed 
out by Levy-Leblond's detailed account\cite{L-L} where the relevant 
references on this matter can be found. It is not strange that the 
recent history of physics had forgotten and considered unexistant the 
$(s\neq0)$ region of CM. The spin is neither relativistic nor a 
quantum mechanical property of the electron. The only quantum 
mechanical aspect of the electron spin is that it is quantized. 

In QM spin is usually defined as the translationally invariant part of 
the total angular momentum operator, or in an equivalent way, as the 
total angular momentum measured in the center of mass frame of the 
particle, when this frame is available. In the description of photons 
we have no center of mass frame but the first definition still works. 
But this definition is related to the representation of the generators 
of the rotation group on the Hilbert space of states. Even more, the 
QM spin dynamics is not postulated, it is contained in its commutation 
relations with the total Hamiltonian. 

However, in CM spin is usually defined by assuming new degrees of 
freedom, the spin degrees of freedom, and even a plausible dynamical 
equation for the spin evolution. Let us mention the case of the 
Bargmann, Michel and Telegdi equation\cite{Bargmann}. 

Following with the above scheme of Fig.~\ref{fig:fig1} we see that 
physics first entered into the QM $(s\neq0)$ region by hand. Remember 
the Pauli method of introducing $\sigma$-spin matrices into 
Schroedinger's equation, or for example when Dirac performs the 
linearization of Klein-Gordon operator to obtain the quantum 
mechanical relativistic description of the electron. But it is after 
Wigner's work\cite{Wigner} on the inhomogeneous Lorentz group 
representations, that by finding the different representations of the 
Poincar\' e group, we can obtain the description of particles of any 
spin. But Wigner's work not only develops the QM of spinning systems 
but it also provides a very precise mathematical definition of the 
concept of elementary particle at the quantum level. An elementary 
particle is defined as a quantum mechanical system whose Hilbert space 
of pure states is the representation space of a projective unitary 
irreducible representation of the Poincar\' e group. 

The very important expression with physical consequences, of the above 
mathematical definition lies in the term {\sl irreducible}. 
Mathematically it means that the Hilbert space is an invariant vector 
space under the group action and that it has no other invariant 
subspaces. But it also means that there are no other states for a 
single particle that those that can be obtained by just taking any 
arbitrary vector state, form all its posible images in the different 
inertial frames and finally produce all linear combinations of these 
vectors. 

We see that starting from a single state and by a simple change of 
inertial observer, we obtain the state of the particle described in 
this new frame. Then let us take the orthogonal part of this vector to 
the previous one and normalize it. Repeat this operation with another 
kinematical transformation acting on the same first state followed by 
the corresponding orthonormalization procedure, as many times as 
neccesary to finally obtain a complete orthonormal basis of the whole 
Hilbert space of states. All states in this basis are characterized by 
the physical parameters that define the first state and a countable 
collection of group transformations of the kinematical group $G$. And 
this can be done starting from any arbitrary state. 

This idea allows us to define a concept of physical equivalence among 
states of any arbitrary quantum mechanical system in the following 
way: Two states are said to be physically equivalent if they can 
produce by the above method an orthonormal basis of the same Hilbert 
subspace, or in an equivalent way, if they belong to the same 
invariant subspace under the group action. It is easy to see that this 
is an equivalence relation. But if the representation is irreducible, 
all states are equivalent as basic pieces of physical information for 
describing the elementary system. There is one and only one single 
piece of basic physical information to describe an elementary object. 
That is what the term elementary means. 

But this definition of elementary particle is a pure group theoretical 
one. The only quantum mechanical ingredient of it is that the group 
operates on a Hilbert space. Then one question arises. Can we 
translate this QM definition into the classical domain and then obtain 
an equivalent group theoretical definition for a classical elementary 
particle? If this is the case we will be able to develop the 
$(s\neq0)$ region of CM in a more equivalent way. The answer is 
affirmative. 

In CM we have no vector space structure to describe the states of a 
system. What we have are manifolds of points where each point 
represents either the configuration state, the kinematical state or 
the phase state of the system depending on which manifold we work. But 
the idea that any point that represents the state of an elementary 
particle is physically equivalent to any other, is in fact the very 
mathematical concept of homogeneity of the manifold under the 
corresponding group action. In this way, the irreducibility assumption 
of the QM definition is translated into the realm of CM in the concept 
of homogeneity of the corresponding manifold under the Poincar\' e 
group or any other kinematical group we consider as the symmetry group 
of the theory. But, what manifold? Configuration space? Phase space? 
The answer as has been shown in previous 
papers\cite{Rivas1},\cite{Rivas2},\cite{Rivas3}, is that the 
appropriate manifold is the {\sl kinematical space}. 

In previous attempts, Bacry\cite{Bacry}, considered the phase space as 
the homogeneous manifold to describe elementary objects arriving at 
the conclusion that the most general elementary particle is a system 
of four degrees of freedom. Three represent the position of the 
particle, being the linear momentum their conjugate variables, and the 
fourth is an angle $\alpha$ whose conjugate momentum is a spin 
component $S_\alpha$ such that the spin components are expressed in 
terms of these two variables and an invariant value $S$, the absolute 
value of the spin, in the form: 
 \[ S_x=(S^2-S_\alpha^2)^{1/2}\cos\alpha,\quad S_y=(S^2-
S_\alpha^2)^{1/2}\sin\alpha,\quad S_z=S_\alpha. 
 \] 
This result was generalized independently by Kirillov, Kostant and 
Souriau and is known as the KKS theorem, showing that the coadjoint 
action of any Lie group defines on its orbits a symplectic 
structure\cite{KKS}. 

But the phase space, although interpreted as the state space of CM 
does not play the same role as the Hilbert space in QM at least as the 
dynamics is concerned. In QM the dynamics is stated in terms of 
initial $|\psi_i>$ and final $|\psi_f>$ states, such that the 
probability amplitude for the dynamical process $|\psi_i>\to |\psi_f>$ 
is given by the corresponding matrix element of the scattering 
operator $<\psi_f|S|\psi_i>$. But both $|\psi_i>$ and $|\psi_f>$ are 
elements of the same Hilbert space that at the same time it plays the 
role of the space that describes all the particle states, it also 
represents the kinematical space where the dynamics is running. 

In the Lagrangian approach of CM the kinematical space is the manifold 
spanned by the initial (or final) boundary variables that are held 
fixed in the corresponding variational formulation. It is this 
manifold $X$ where the dynamics is developed, where when quantizing 
the system we obtain the natural link between the classical and 
quantum formalisms through the Feynman's path integral approach, as 
shown in \cite{Rivas3}. It is this manifold the natural base space to 
define on it the Hilbert space structure of the quantized system. In a 
formal way we can say that each point $x\in X$, of the kinematical 
space of the Lagrangian formalism that represents the inital or final 
kinematical state, smears out and is transformed through the Feynman's 
quantization into the squared integrable function $\psi(x)$ with 
support around that point $x$ and representing the particle wave 
function. 

Usually the Lagrangian of any classical system is restricted to depend 
only on the first order derivative of each of the variables $q_i$ that 
represent the different independent degrees of freedom, or 
equivalently, that the $q_i$ satisfy second order differential 
equations. But at this stage, if we do not know what are the basic 
variables we need to describe our elementary system how can we state 
that they necessarily satisfy second order differential equations? If 
some of the degrees of freedom, say $q_1$, $q_2$ and $q_3$, represent 
the center of mass position of the system, Newtonian mechanics implies 
that in this particular case $L$ will depend on the first order 
derivatives of these three variables. But what about other degrees of 
freedom? It is the homogenenity condition on the kinematical space, as 
the mathematical statement of elementarity, that will restrict the 
dependence of the Lagrangian on these higher order derivatives. It is 
the definition of elementary particle that will suply the structure of 
the Lagrangian. 

A classical elementary particle is defined as a Lagrangian classical 
system whose kinematical space $X$, is a homogeneous space of the 
corresponding kinematical group $G$. \cite{Rivas1,Rivas2,Rivas3} 

Since the Galilei and Poincar\' e groups are 10-parameter Lie groups, 
the largest homogeneous space we can find is a 10-dimensional manifold 
whose variables share the same domain and dimensions like the 
variables we use to parameterize the group. Both groups are 
parameterized in terms of the following variables $(b,{\bf a},{\bf 
v},{\bf\alpha})$ with domains and dimensions respectively like $b\in R$ 
that represents the time parameter of the time translation, ${\bf a}\in 
R^3$ the three spacial coordinates for the space translation, ${\bf 
v}\in R^3$ the three components of the relative velocity of the 
inertial observers, that are restricted to $v<c$ in the Poincar\' e 
case, and finally ${\bf\alpha}\in SO(3)$ are three angular variables 
that characterize the relative orientation of the corresponding 
Cartessian frames and whose compact domain is expressed in terms of a 
suitable parametrization of the rotation group. 

In this way the maximum number of kinematical variables is also 10, 
and we represent them by $x\equiv(t,{\bf r},{\bf u},{\bf\alpha})$ with 
the same domains and dimensions as above and are interpreted 
respectively as the time, position, velocity and orientation of the 
particle. This is the same description of the initial and final states 
as the relativistic spherical top in the Hanson and Regge canonical 
approach\cite{Hanson}, but in that work different constraints are used 
to avoid dependence of the Lagrangian on second order derivatives. 

In our approach, since the Lagrangian must also depend on the next order 
derivatives of the kinematical variables we arrive at the conclusion 
that $L$ must also depend on the acceleration and angular velocity of 
the particle. The particle is a system of six degrees of freedom, 
three, ${\bf r}$, represent the position of a point and other three 
${\bf\alpha}$, its orientation in space, which we can visualize by 
assuming a system of three orthogonal unit vectors linked to point 
${\bf r}$ as a body frame, but the Lagrangian may depend up to the 
second time derivative of ${\bf r}$, or acceleration of that point, 
and on the first derivative of ${\bf\alpha}$, i.e., the angular 
velocity. By this definition it is the kinematical group $G$ that 
represents the special Relativity Principle that completely determines 
the structure of the Lagrangian that represents a classical elementary 
particle\cite{Rivas1,Rivas2,Rivas3}. Point particles are particular 
cases of the above definition and their kinematical spaces are just 
the quotient structures between the group $G$ and the homogeneous 
transformations subgroup of rotations and boosts, and thus their 
kinematical variables reduce only to time and position $(t,{\bf r})$. 

We began this section by considering the classical limit of QM and we 
have remarked the different treatment the recent history of physics 
has done to the $(s\neq0)$ region in both QM and CM formalisms. But 
this difference between the two approaches might have physical 
consequences. For instance, let us consider the well known QM 
diffraction experiment of a beam of particles by a certain slit. If we 
send a beam of point particles we expect in the screen a replica of 
the slit shape and size but if we obtain a diffraction pattern we 
interpret this outcome as a pure quantum mechanical effect. This 
interpretation is correct if for instance we perform the experiment 
with pions, i.e., spinless particles. But if the experiment is done 
with photons, electrons, neutrons, etc. as is usually the case, all of 
them are spinning particles. Since the spin of these particles is of 
order of $\hbar$ not the whole outcome of the experiment of order 
$\hbar$ is a quantum mechanical effect, because part of it must be 
related to the spin structure of the particles and must be analyzed 
theoretically. What is a quantum mechanical effect is the difference 
between the real outcome and that part that can be interpreted in 
classical terms but taking into account the corresponding classical 
spin contribution. A deeper understanding of the classical spin 
description will help to better understand the quantum mechanical 
effects. 

In this work, and following with the above argument we shall consider 
the classical contribution to the tunnel effect. Classically, a point 
particle confined into a potential well can never escape from it if 
its energy is less than the top of the potential barrier. In QM there 
is a nonvanishing probability of crossing. However, if the particle 
has spin and this spin is of orbital nature as will be described in 
the next section, we shall see in Section V that because of the spin, 
the classical spinning particle can cross the barrier even with 
kinetic energies below the potential energy barrier, and that this 
contribution is not in general negligible when compared with the 
quantum tunnel effect.

\section{Classical Particle with (anti)-orbital spin}
\label{sec:particle}

We are going to analyze in detail a non-relativistic Lagrangian that 
although it does not give rise to a spin $1/2$ particle when quantized 
(see \cite{Rivas3}), however it behaves in most of the features 
simmilarly as the relativistic electron described in the mentioned 
reference, but the mathematics involved are simpler than in the 
general case, and illustrates the spin structure and dynamics for the 
example we want to work out in section V. 

Let ${\cal G}$ be the Galilei group. Let us consider a Galilei 
particle whose kinematical space is $X={\cal G}/SO(3)$, so that any 
point $x\in X$ can be characterized by the seven variables 
$x\equiv(t,{\bf r},{\bf u})$, ${\bf u}=d{\bf r}/dt$, which are 
interpreted as the time, position and velocity of the particle 
respectively. In this example we have no orientation variables and 
because of this we cannot have, when quantizing the system, half 
integer spin values. The Lagrangian will also depend on the next order 
derivatives, i.e., on the acceleration of the particle. Rotation and 
translation invariance implies that $L$ will be a function of only 
${\bf u}^2$, $(d{\bf u}/dt)^2$ and ${\bf u}\cdot d{\bf u}/dt= 
d(u^2/2)/dt$, but being this last term a total derivative it will not 
be considered here. 

Let us assume that our elementary system is represented by the 
following Lagrangian 
 \begin{equation}
L={m\over2}\left({d{\bf r}\over dt}\right)^2-
{m\over2\omega^2}\left({d^2{\bf r} \over dt^2}\right)^2,
\label{eqn1}
 \end{equation} 
where $m$ is the mass and the parameter $\omega$ of dimensions of 
s$^{-1}$ represents an internal frequency. In terms of the kinematical 
variables and their derivatives, and in terms of some group invariant 
evolution parameter $\tau$, the Lagrangian can be written as 
 \begin{equation}
L={m\over2}{\dot{\bf r}^2\over\dot 
t}-{m\over2\omega^2}{\dot{\bf u}^2 \over\dot t},
\label{eqn2}
 \end{equation} 
where the dot means $\tau-$derivative. If we consider that the 
evolution parameter is dimensionless, all terms in the Lagrangian have 
dimensions of action. Because the Lagrangian is a homogeneous function 
of first degree in terms of the derivatives of the kinematical 
variables, $L$ can also be written as 
 \begin{equation}
L=-T\dot t+{\bf R}\cdot\dot{\bf r}+{\bf U}\cdot\dot{\bf u},
\label{eqn3}
 \end{equation} 
where the functions accompanying the derivatives of the kinematical 
variables are defined and explicitely given by 
 \begin{eqnarray} 
T&=&-{\partial L\over\partial\dot 
t}={m\over2}\left({d{\bf r}\over dt}\right)^2-
{m\over2\omega^2}\left({d^2{\bf r} \over dt^2}\right)^2,\nonumber\\ 
{\bf R}&=&{\partial L\over\partial\dot{\bf r}}=m{d{\bf r}\over 
dt},\label{eqn4}\\ {\bf U}&=&{\partial L\over\partial\dot{\bf u}}=-
{m\over\omega^2}{d^2{\bf r}\over dt^2}.\nonumber 
 \end{eqnarray}

Dynamical equations are: 
 \begin{equation}
{1\over\omega^2}{d^4{\bf r}\over 
dt^4}+{d^2{\bf r}\over dt^2}=0,
\label{eqn5}
 \end{equation} 
whose general solution is: 
 \begin{equation} 
{\bf r}(t)={\bf A}+{\bf B}t+{\bf C}\cos\omega t+{\bf 
D}\sin\omega t, \label{eqn6} 
 \end{equation} 
in terms of the 12 integration constants ${\bf A}$, 
${\bf B}$, ${\bf C}$ and ${\bf D}$. 

When applying Noether's theorem to the invariance of dynamical 
equations under the Galilei group, the corresponding constants of the 
motion can be written in terms of the above functions in the form: 

 \begin{eqnarray} 
\hbox{\rm Energy}\quad H&=&T-{\bf u}\cdot{d{\bf 
U}\over dt}= {m\over2}{\bf u}^2-{\omega^2\over2m}{\bf U}^2-{\bf 
u}\cdot{d{\bf U}\over dt},\label{eqn71}\\ \hbox{\rm Linear 
Momentum}\quad {\bf P}&=&{\bf R}-{d{\bf U}\over dt}=m{\bf u}-{d{\bf 
U}\over dt},\label{eqn72}\\ \hbox{\rm Galilei or Static Momentum}\quad 
{\bf K}&=&m{\bf r}-{\bf P}t-{\bf U},\label{eqn73}\\ \hbox{\rm Angular 
Momentum}\quad {\bf J}&=&{\bf r}\times{\bf P}+{\bf u}\times{\bf 
U}={\bf L}+{\bf S}.\label{eqn7} 
 \end{eqnarray} 
It is the presence of the ${\bf U}$ function that distinguishes the 
features of this system with respect to the point particle case. We 
see that the total linear momentum is not lying along the direction of 
the velocity ${\bf u}$. The translationally invariant part of the 
total angular momentum ${\bf J}$ is the spin of the system ${\bf S}$. 
It is the dependence on the acceleration that in this system of three 
degrees of freedom leads to the spin structure, such that the spin 
takes the form: 
 \begin{equation} 
{\bf S}={\bf u}\times{\bf 
U}={m\over\omega^2}\;{d^2{\bf r}\over dt^2}\times {d{\bf r}\over dt},
\label{eqn8}
 \end{equation} 
is always orthogonal to the velocity and acceleration 
of point ${\bf r}$, and its dynamics is given by taking the time 
derivative of ${\bf J}$ in (\ref{eqn7}) and taking into account that 
${\bf P}$ is constant it leads to 
 \begin{equation}
{d{\bf S}\over dt}={\bf P}\times{\bf u}.
\label{eqn9}
 \end{equation} 
The spin is not a constant of the motion except in the 
center of mass frame, it precesses along the linear momentum ${\bf P}$ 
such that its projection ${\bf S}\cdot{\bf P}$, called helicity, 
remains constant. 

If we substitute the general solution (\ref{eqn6}) in 
(\ref{eqn71}-\ref{eqn7}) we see in fact that 
the integration constants are related to the above conserved 
quantities \begin{eqnarray*} H&=&{m\over2}{\bf B}^2-
{m\omega^2\over2}({\bf C}^2+{\bf D}^2),\\ {\bf P}&=&m{\bf B},\\ {\bf 
K}&=&m{\bf A},\\ {\bf J}&=&{\bf A}\times m{\bf B}-m\omega{\bf 
C}\times{\bf D}. \end{eqnarray*} 

But at the same time that the particle acquires its spin we see 
that the static momentum ${\bf K}$ differs from the point particle 
case in the term $-{\bf U}$, such that if we define the vector ${\bf 
x}={\bf U}/m$, then $\dot{\bf K}=0$ leads from (\ref{eqn73}) to the 
equation: 
 \[
{\bf P}=m{d({\bf r}-{\bf x})\over dt},
 \] 
and ${\bf q}={\bf r}-{\bf x}$, defines the position of the center 
of mass of the particle that is a different point than ${\bf r}$ and 
is given by 
 \begin{equation}
{\bf q}={\bf r}+{1\over\omega^2}\;{d^2{\bf r}\over dt^2},
\label{eqn10}
 \end{equation} 
such that in terms of it, dynamical equations can be 
separated into the form: 
\begin{eqnarray} {d^2{\bf q}\over 
dt^2}&=&0,\label{eqn110}\\ {d^2{\bf r}\over dt^2}&+&\omega^2({\bf r}-
{\bf q})=0.\label{eqn111} 
 \end{eqnarray}

In terms of these variables the spin takes the form
 \begin{equation}
{\bf S}=-({\bf r}-{\bf q})\times m{\bf u},
\label{eqn12}
 \end{equation} 
and it has the formal expresion of the opposite of an 
orbital angular momentum with respect to point ${\bf q}$ of a point 
mass $m$ at point ${\bf r}$ moving at the speed ${\bf u}$. We say that 
this spin is of (anti)-orbital nature. 

Point ${\bf q}$ moves in a straight trajectory at constant velocity 
while the motion of point ${\bf r}$ is an isotropic harmonic 
oscillator of angular frequency $\omega$ around point ${\bf q}$. But 
if ${\bf q}$ represents the center of mass position then, what 
position does point ${\bf r}$ represent? Point ${\bf r}$ represents 
the charge of the particle position. This can be seen by considering 
some interaction with an external field. The homogenenity condition of 
the Lagrangian in terms of the derivatives of the kinematical 
variables leads us to consider an interaction term of the form 
 \begin{equation}
L_I=-e\phi(t,{\bf r})\dot 
t+e{\bf A}(t,{\bf r})\cdot\dot{\bf r},
\label{eqn13}
 \end{equation} 
linear in the derivatives and where the external 
potentials are functions of $t$ and ${\bf r}$. We can also consider 
more general interaction terms depending on ${\bf u}$ and $\dot{\bf 
u}$, but this will not be an interaction with an external 
electromagnetic field. Then the dynamical equations are 
 \begin{equation} {1\over\omega^2}{d^4{\bf r}\over dt^4}+{d^2{\bf 
r}\over dt^2}={e\over m} \left({\bf E}(t,{\bf r})+{\bf u}\times{\bf 
B}(t,{\bf r})\right), \label{eqn14} 
 \end{equation} where the electric ${\bf E}$ and magnetic field ${\bf 
B}$ are expressed in terms of the potentials in the usual form, and 
because the interaction term does not affect the dependence of the 
Lagrangian on $\dot{\bf u}$, the spin and the center of mass 
definitions (\ref{eqn8}) and (\ref{eqn10}) respectively, remain the 
same as in the free particle case. Then dynamical equations can again 
be separated in the form 
 \begin{eqnarray} {d^2{\bf q}\over dt^2}&=&{e\over m}\left({\bf 
E}(t,{\bf r})+{\bf u}\times{\bf B}(t,{\bf r})\right),\label{eqn151}\\ 
{d^2{\bf r}\over dt^2}&+&\omega^2({\bf r}-{\bf q})=0,\label{eqn152} 
 \end{eqnarray} where the center of mass ${\bf q}$ satisfies Newton's 
equations under the action of the total external Lorentz force, while 
point ${\bf r}$ still satisfies the isotropic harmonic motion of 
angular frequency $\omega$ around point ${\bf q}$. This internal 
motion remains unaffected by the external interaction. But the 
external force and the fields are defined at point ${\bf r}$ and not 
at point ${\bf q}$. Point ${\bf r}$ clearly represents the point 
charge position. In fact, this minimal coupling we have considered is 
the coupling of the electromagnetic potentials with the particle 
current, that in the relativistic case can be written as $A^\mu 
j_\mu$, but the current $j_\mu$ is associated to the motion of point 
${\bf r}$. This charge has an oscillatory motion of very high 
frequency $\omega$ that in the case of the relativistic electron (see 
\cite{Rivas2,Rivas3}), is $\omega=2mc^2/\hbar\simeq 10^{20}$s$^{-1}$. The 
average position of the charge is the center of mass, but it is its 
internal rotational motion, usually known as the zitterbewegung, that 
gives rise to the spin structure and to the magnetic properties of the 
particle. 

When analyzed in the center of mass observer (see 
Fig.~\ref{fig:fig2}), the system reduces to a point charge whose 
motion is in general an ellipse but if we choose $C=D$, and ${\bf 
C}\cdot{\bf D}=0$, it reduces to a circle of radius $a=C=D$, 
orthogonal to the spin that is constant in this frame. Then if the 
particle has charge $e$, it has a constant magnetic moment 
 \[
{\bf\mu}={1\over 2}{\bf 
r}\times{\bf j}={e\over 2}{\bf r}\times{\bf u}= -{e\over 2m}{\bf S},
 \] 
opposite to the spin direction if $e>0$, and a nonvanishing 
oscilating electric dipole ${\bf p}=e{\bf r}$, orthogonal to both 
${\bf\mu}$ and ${\bf S}$ in this frame, such that its average value 
vanishes for times larger than the natural period of this internal 
motion.

In this case the above constants of the motion are no longer conserved 
quantities. For instance, the linear momentum satisfies 
 \[
{d{\bf P}\over 
dt}={\bf F},
 \] where ${\bf F}$ is the external Lorentz force. If we take the time 
derivative of the total angular momentum it leads to 
 \begin{equation} {d{\bf J}\over dt}={\bf r}\times{d{\bf P}\over 
dt}+{d{\bf r}\over dt} \times{\bf P}+{d{\bf S}\over dt}. \label{eqnJ} 
 \end{equation} 
The spin dynamics is given by 
 \begin{equation} {d{\bf S}\over dt}={m\over\omega^2}{d^3{\bf r}\over 
dt^3}\times{d{\bf r}\over dt} ={\bf P}\times{\bf u}, 
 \end{equation} and remains the same expression as in the free case 
(\ref{eqn9}), but now ${\bf P}$ is not a constant function. If we 
substitute this in the previous expression (\ref{eqnJ}), it turns out 
that the total angular momentum satisfies 
 \begin{equation}
{d{\bf J}\over dt}={\bf r}\times{\bf F},
\label{eqn16}
 \end{equation} where on the right hand side we have the total 
external torque produced by the fields. Although the system has 
magnetic moment associated to the charge motion, because of the 
minimal coupling interaction, the external electromagnetic field 
produces forces, and the corresponding torques with respect to the 
origin, but no other torques, so that a spin evolution equation of the 
form 
 \[
{d{\bf S}\over 
dt}={\bf\mu}\times{\bf B},
 \] as in Ref.~\cite{Bargmann}, does not arise in this model in which 
the electromagnetic structure of the particle is just that of a point 
charge without any intrinsic magnetic moment, as it happens to be the 
case of the relativistic electron, where the charge motion in the 
center of mass frame is a circle of radius $a=S/mc=\hbar/2mc$ at the 
speed of light $c$. Only for anomalous magnetic moments we should have 
to consider the addition of external torques of the above form, where 
${\bf\mu}$ will be the anomalous magnetic moment. But in that case we 
need to have in the interaction Lagrangian terms of the form 
$F^{\mu\nu}M_{\mu\nu}/2$, where $M_{\mu\nu}$ is the dipole tensor that 
describes the intrinsic dipole structure of the particle. 

If we consider that the mechanical energy of the particle is still 
given by expression (\ref{eqn71}), then because of the interaction it 
changes in the form 
 \[
{dH\over dt}={\bf u}\cdot{\bf F}=e{\bf u}\cdot{\bf E}.
 \] The change of the mechanical energy of the particle is the work of 
the external force along the charge trajectory. If the electric field 
is conservative then the change of this energy between two arbitrary 
points of the trajectory is 
 \begin{eqnarray*} H(t_2)&-&H(t_1)=e\phi(t_1,{\bf r}_1)-e\phi(t_2,{\bf 
r}_2)\\ =e\left(\phi(t_1,{\bf q}_1)-\phi(t_2,{\bf 
q}_2)\right)&+&e\left(\phi(t_1,{\bf r}_1)-\phi(t_1,{\bf q}_1)\right)+ 
e\left(\phi(t_2,{\bf q}_2)-\phi(t_2,{\bf r}_2)\right)\\ \simeq 
e\phi(t_1,{\bf q}_1)-e\phi(t_2,{\bf q}_2),&& 
 \end{eqnarray*} if the difference of potential between the center of 
mass and center of charge position at any instant can be considered 
negligible if the external field is of smooth variation. 

Once we introduce the classical path into the free Lagrangian and 
integrate from the initial to the final point, we obtain the action 
function for the free system along this path that can be written in 
terms of the boundary kinematical variables in the form 
\begin{eqnarray*} A(t_1,{\bf r}_1,{\bf u}_1;t_2,{\bf r}_2,{\bf 
u}_2)&=&{m\over2\Delta}\bigg[ {1\over\omega}({\bf u}_2-{\bf 
u}_1)^2\sin\omega(t_2-t_1)\\ &+&\omega({\bf r}_2-{\bf 
r}_1)^2\sin\omega(t_2-t_1)\\ &-&(t_2-t_1)\,({\bf u}_2-{\bf 
u}_1)^2\cos\omega(t_2-t_1)\\ &+&2(t_2-t_1)\,({\bf u}_2\cdot{\bf 
u}_1)\,(1-\cos\omega(t_2-t_1))\\ &-& 2({\bf r}_2-{\bf r}_1)\cdot({\bf 
u}_1+{\bf u}_2)\,(1-\cos\omega(t_2-t_1))\bigg], \end{eqnarray*} where 
$\Delta=2(1-\cos\omega(t_2-t_1))-\omega(t_2-t_1)\sin\omega(t_2-t_1)$. 
This is the action function that must be considered when analyzing 
Feynman's kernel in the path integral approach of this system.

\section{Canonical Formalism}
\label{sec:canonical}

Although the Lagrangian depends on second order derivatives we can 
develop the corresponding canonical formalism. Starting from the 
Lagrangian 
 \[
L={m\over2}\,\dot{\bf r}^2-{m\over2\omega^2}\,\ddot{\bf r}^2,
 \] where now the dot means time derivative, we have in this case six 
generalized coordinates that are not the independent degrees of 
freedom but they are the three degrees of freedom ${\bf q}_1={\bf r}$ 
and their first derivatives ${\bf q}_2= d{\bf r}/dt$, such that the 
conjugate momenta are defined by\cite{Whittaker} 
 \[ {\bf p}_1={\partial L\over\partial\dot{\bf r}}-{d\over 
dt}\left({\partial L\over\partial\ddot{\bf r}}\right)={\bf R}-{d{\bf 
U}\over dt},\quad {\bf p}_2={\partial L\over\partial\ddot{\bf r}}={\bf 
U}. 
 \] The phase space is a 12-dimensional manifold and the Hamiltonian 
is in fact the total energy written in terms of the canonical 
variables 
 \[ H={\bf p}_1\cdot\dot{\bf q}_1+{\bf p}_2\cdot\dot{\bf q}_2-L= {\bf 
p}_1\cdot{\bf q}_2-{m\over2}\,{\bf q}_2^2-{\omega^2\over2m}\,{\bf 
p}_2^2. 
 \]
Hamilton-Jacobi equations are
 \[ \dot{\bf q}_1={\partial H\over\partial{\bf p}_1}={\bf q}_2,\quad 
\dot{\bf p}_1=-{\partial H\over\partial{\bf q}_1}=0, 
 \]
 \[ \dot{\bf q}_2={\partial H\over\partial{\bf p}_2}=-{\omega^2\over 
m}{\bf p}_2,\quad \dot{\bf p}_2=-{\partial H\over\partial{\bf q}_2}=-
{\bf p}_1+m{\bf q}_2. 
 \]

The ten Noether constants of the motion become in this formalism the 
generating functions of the corresponding canonical transformations of 
time and space translations, pure Galilei transformations and 
rotations and are explicitely given by \begin{eqnarray*} H&=&{\bf 
p}_1\cdot{\bf q}_2-{m\over2}\,{\bf q}_2^2-{\omega^2\over2m}\,{\bf 
p}_2^2,\\ {\bf P}&=&{\bf p}_1,\\ {\bf K}&=&m{\bf q}_1-{\bf p}_1t-{\bf 
p}_2,\\ {\bf J}&=&{\bf q}_1\times{\bf p}_1+{\bf q}_2\times{\bf p}_2, 
\end{eqnarray*} and since the Poisson bracket of two constants of the 
motion is again a constant of the motion we obtain that the above 
constants of the motion satisfy the commutation relations 
 \[ [J_i,J_k]=\epsilon_{ikl}J_l,\quad 
[J_i,P_k]=\epsilon_{ikl}P_l,\quad [J_i,K_k]=\epsilon_{ikl}K_l, \quad 
[J_i,H]=0, 
 \]
 \[ [P_i,P_k]=0,\quad [P_i,H]=0,\quad [K_i,K_k]=0,\quad 
[K_i,H]=P_i,\quad [K_i,P_j]=m\delta_{ij}, 
 \] where $[.,.]$ is the corresponding Poisson bracket and because 
$[K_i,P_j]\neq 0$ they are not the commutations relations of the 
Galilei group but rather those of the extended Galilei group\cite
{L-LGalileo}. 
Although $K_i$ satisfies $[K_i,H]=P_i$, it is a constant of the motion 
because as we see in (\ref{eqn73}) that ${\bf K}$ is time dependent 
its total time derivative is 
 \[ {d{\bf K}\over dt}=[{\bf K},H]+{\partial{\bf K}\over\partial 
t}={\bf p}_1 -{\bf p}_1=0. 
 \]

The spin observable ${\bf S}={\bf u}\times{\bf U}={\bf q}_2\times{\bf 
p}_2$ satisfies the commutation relations 
 \[ [S_i,S_k]=\epsilon_{ikl}S_l,\quad 
[J_i,S_k]=\epsilon_{ikl}S_l,\quad [S_i,P_k]=0, 
 \]
 \[ [S_i,K_j]=-\epsilon_{ijk}U_k=-\epsilon_{ijk}p_{2k},\quad 
[S_i,H]={dS_i\over dt} =({\bf P}\times{\bf u})_i, 
 \] showing respectively that it is an angular momentum, that 
transforms like a vector under rotations, it is invariant under space 
translations but not under pure Galilei transformations and is not a 
constant of the motion, but satisfies the dynamical equations 
(\ref{eqn9}). We can check that the two constants of the motion 
 \begin{equation}
E=H-{{\bf P}^2\over 2m},\quad \hbox{\rm and}\quad Z^2=\left({\bf J}-
{1\over m}{\bf K}\times{\bf P}\right)^2,
\label{eqncas}
 \end{equation} commute with the above ten generators and they are 
Galilei invariant properties of the particle, that together with the 
mass $m$ completely characterize the structure of this particle. They 
are the three functionally independent Casimir invariants of the 
extended Galilei group. In fact $E$ is the Galilei internal energy of 
the particle and $Z^2$ is the squared of the total angular momentum or 
spin for the center of mass observer\cite{L-LGalileo}.

\section{Classical Tunnel Effect}
\label{sec:tunel}

Let us consider a spinning particle as described above under the 
influence of a potential barrier. Sharp walls correspond classically 
to infinite forces so that we shall consider potentials that give rise 
to finite forces like those of the shape depicted in 
Fig.~\ref{fig:fig3}, where $V_0$ represents the top of the potential. 
Then the external force $F(x)$, is constant and directed leftwards in 
the region $x\in (-a,0)$ and directed rightwards for $x\in (0,b)$, 
vanishing outside these regions. 

Let us assume for simplicity that the spin is pointing up or down in 
the $z$ direction such that the point charge motion takes place in the 
$XOY$ plane. Let $q_x$, $q_y$ and $q_z=0$, be the coordinates of the 
center of mass and $x$, $y$ and $z=0$, the position of the charge. 

Dynamical equations are
 \begin{equation}
{d^2q_x\over dt^2}={1\over m}F(x),\quad{d^2q_y\over dt^2}=0,
 \end{equation}
 \begin{equation}
{d^2x\over dt^2}+\omega^2(x-q_x)=0,\quad{d^2y\over dt^2}+\omega^2(y-
q_y)=0,
 \end{equation}
where
\[
F(x)=\left\{\begin{array}{ll}
-{eV_0/a},&\mbox{for $x\in(-a,0)$},\\
{eV_0/b},&\mbox{for $x\in(0,b)$},\\
0, &\mbox{otherwise}.
\end{array}\right.
 \]

To make the corresponding numerical analysis we shall define different 
dimensionless variables. Let $R$ be the average separation between the 
center of charge and center of mass. Then we define the new 
dimensionless position variables: 
 \[ \hat q_x=q_x/R,\quad \hat q_y=q_y/R,\quad \hat x=x/R,\quad \hat 
y=y/R,\quad \hat a=a/R,\quad \hat b=b/R. 
 \]

The new dimensionless time variable $\alpha=\omega t$ is just the 
phase of the internal motion, such that dynamical equations become 
 \[
{d^2\hat q_x\over d\alpha^2}=A(\hat x),\quad{d^2\hat q_y\over 
d\alpha^2}=0,
 \]
 \[
{d^2\hat x\over d\alpha^2}+\hat x-\hat q_x=0,\quad{d^2\hat y\over 
d\alpha^2}+\hat y-\hat q_y=0,
 \]
where $A(x)$ is given by
 \[
A(\hat x)=\left\{\begin{array}{ll}
-{eV_0/\hat am\omega^2R^2},&\mbox{for $\hat x\in(-\hat a,0)$},\\
{eV_0/\hat bm\omega^2R^2},&\mbox{for $\hat x\in(0,\hat b)$},\\
0, &\mbox{otherwise}.
\end{array}\right.
 \]

In the case of the electron, the internal velocity of the charge is 
$\omega R=c$, (see \cite{Rivas2} and \cite{Rivas3}), so that the 
parameter $e/mc^2=1.9569\times10^{-6}$V$^{-1}$, such that for 
potentials of order of 100 volts we can take the dimensionless parameter 
$eV_0/m\omega^2R^2\simeq2\times10^{-4}$. 

If we choose as initial conditions for the center of mass motion 
 \[
\hat q_y(0)=0,\quad d\hat q_y(0)/d\alpha=0,
 \] then the center of mass is moving along $OX$ axis. Then the above 
system reduces to the analysis of the one-dimensional motion where the 
only variables are $\hat q_x$ and $\hat x$. Let us call from now on 
these variables $q$ and $x$ respectively and remove all hats from the 
dimensionless variables. Then the dynamical equations to be solved 
numerically are just 
 \begin{equation}
{d^2 q\over d\alpha^2}=A(x),\quad {d^2 x\over 
d\alpha^2}+x-q=0,
 \end{equation} 
where $A(x)$ is given by 
\begin{equation}A(\hat x)=\left\{
\begin{array}{ll}
-2a^{-1}10^{-4},&\mbox{for $x\in(-a,0)$},\\
2b^{-1}10^{-4},&\mbox{for $x\in(0,b)$},\\
0,&\mbox{otherwise}.
\end{array}\right.
 \end{equation}

Numerical integration has been performed by means of the computer 
package {\sl ODE Workbench}\cite{JMA}. The quality of the numerical 
results is tested by using the different integration schemes this 
program allows, ranging from the very stable embedded Runge-Kutta code 
of eight order due to Dormand and Prince to very fast extrapolation 
routines. All codes have adaptive step size control and we check that 
smaller tolerances do not change the results. 

With $a=b=1$, and in energy units such that the top of the barrier is 
1, if we take an initial kinetic energy $K$ below this threshold, 
$K=m\dot q(0)^2/2eV_0=0.41$ we obtain for the center of mass motion 
the graphic depicted in Fig.~\ref{fig:fig4} where it is shown the 
variation of the kinetic energy of the particle $K(q)$, with the 
center of mass position during the crossing of the barrier. There is 
always crossing with a kinetic energy above this value. In 
Fig.~\ref{fig:fig5} the same graphical evolution with $a=1$ and $b=10$ 
and $K=0.8055$ for a potential of $10^3$ Volts in which the different 
stages in the evolution are  more evident. Below the initial 
values for the kinetic energy of $0.4$ and $0.80$ respectively, the 
particle does not cross the potential barrier and it is rejected 
backwards. 

If in both examples the parameter $a$ is ranged from 1 to 0.05, thus 
making the left slope sharper, there is no appreciable change in the 
crossing energy, so that with $a=1$ held fixed we can compute the 
minimum crossing energies for different $b$ values, $K_c(b)$. Initial 
center of mass position has been ranged from $q(0)=-3.5$ to $-3.0$, 
while the center of charge initial position and velocity from $x(0)=-
2.5$ to $-2.0$, and $\dot x(0)=\dot q(0)$, respectively. 

To compare this model with the quantum tunnel effect, let us consider 
a point particle of mass $m$ and charge $e$, under the same one-dimensional 
potential depicted in Fig.~\ref{fig:fig3}. If the particle 
of total energy $E$, is initially on the left hand side of the 
barrier, the wave function of this system is: 

\begin{equation}\psi(x)=\left\{
\begin{array}{ll}
e^{ikx}+Re^{-ikx},&\mbox{for $x\le -a$},\\
C_1{\rm Ai}(D(1-G+x/a))+C_2{\rm Bi}(D(1-G+x/a)),&\mbox{for $-a\le x\le 0 $},\\
C_3{\rm Ai}(L(1-G-x/b))+C_4{\rm Bi}(L(1-G-x/b)),&\mbox{for $0\le x\le b $},\\
Te^{ikx},&\mbox{for $x\ge b$},\\
\end{array}\right.
 \end{equation}
where $x$ is the same dimensionless position variable as before, and the 
constants
 \begin{equation} 
k=\left(\frac{E}{2mc^2}\right)^{1/2},\quad D=\left(\frac{eV_0a^2}{2mc^2}\right)^{1/3},\quad 
L=\left(\frac{eV_0b^2}{2mc^2}\right)^{1/3},\quad G=\frac{E}{eV_0},$$ 
 \end{equation} 
and ${\rm Ai}(x)$ and ${\rm Bi}(x)$ are the Airy functions of $x$.
The six integration constants $R$, $T$, and $C_i, i=1,2,3,4$, can be 
obtained by assuming continuity of the functions and their first order 
derivatives at the separation points of the different regions. The 
coefficient $|R|^2$ represents the probability of the particle to be 
reflected by the potential and  $|T|^2$ its probability of crossing.

Computed the $T$ variable for $a=1$ and different values of the 
potential width $b$, and for energies below the top of the barrier 
$eV_0$ we show on Fig.~\ref{fig:fig6} the average probability for 
quantum tunneling for four different extraction potentials $V_0$ of 
$10^2$, $10^3$, $10^4$ and $10^5$ Volts. This average probabibility has been 
computed by assuming that on the left of the barrier there is a 
uniform distribution of particles of energies below $eV_0$. 

Let us consider for the classical spinning particle the same 
uniform distribution of particles. Then, the function $P(b)=1-K_c(b)$, 
where $K_c(b)$ is the minimum kinetic energy for crossing computed 
before, represents the ratio of the electrons that with kinetic energy 
below the top of the potential, cross the barrier because of the spin 
contribution. This function is also depicted in Fig.~\ref{fig:fig6}, 
and clearly has a decreasing behaviour with the distance $b$. The 
different barriers we have considered so far, can be interpreted as 
the one-dimensional extraction potentials of a probe tip placed at a 
variable distance $b$ and at a constant difference of potential $V_0$ 
with respect to the sample. We see that for the different potentials 
shown in that figure this probability of crossing is greater in the 
quantum case than in the classical spin contribution, but they are 
geting closer for higher potentials. 

The main feature of this model that contributes to tunneling is that 
while the center of mass is still on the left hand part of the barrier and 
since the cyclic internal motion of the charge is unchanged by the 
interaction this implies that once the charge penetrates into the left 
slope of the potential during a fraction of its cycle the kinetic 
energy is decreased, while during the remaining time of its cycle it 
is unaffected by external forces so that the center of mass motion is 
a kind of short decceleration periods under a constant force with 
short periods of constant velocity in between as is clearly visible in 
the saw-teeth shape of the particle kinetic energy of 
Fig.~\ref{fig:fig5}. It turns out that the average deccelerating force 
acting on the center of mass of the particle is smaller than in the 
case of a pure point particle. When the particle has completely 
penetrated into the potential and not all its kinetic energy has been 
exhausted, then if the penetration is sufficient such that the charge 
position can reach during part of its cycle the right slope of the 
potential, it turns out that during that time the center of mass is 
under a force directed to the right and thus the kinetic energy is 
increased. It is clear that at least one of the regions $(-a,0)$ or 
$(0,b)$ must be of the order of the radius of the internal motion to 
have an appreciable crossing. In the case of the electron model (see 
\cite{Rivas2} and \cite{Rivas3}) this radius is $\hbar/2mc$, i.e., 
half the Compton wave-length of the electron, so that with $a$ or $b$ 
ot this order of magnitude as has been considered in the above 
examples, we can have a nonvanishing crossing. 

We see that the separation between the center of mass and center of 
charge that gives rise to the spin structure of this particle model 
justifies that this system can cross a potential barrier even if its 
kinetic energy is below the top of the potential.

\acknowledgments

The author is greatly acknowledged to J.M. Aguirregabiria for helpful 
discussions and also for allowing him to use also the Windows beta-version 
of his excellent {\sl ODE Workbench} computer program\cite{JMA}, with 
which most numerical integrations have been performed. This work has 
been partially supported by the Universidad del Pa\'{\i}s Vasco/Euskal 
Herriko Unibertsitatea under contract UPV/EHU 172.310 EB036/95.

\newpage

\begin{figure} \protect\caption{Classical limit of Quantum Mechanics.} 
\label{fig:fig1} \end{figure} 

\begin{figure} \protect\caption{Charge motion in the Center of Mass 
frame.} \label{fig:fig2} \end{figure} 

\begin{figure} \protect\caption{Potential barrier to be crossed by 
spinning particles.} \label{fig:fig3} \end{figure} 
 
\begin{figure} \protect\caption{Evolution of the kinetic energy 
during the crossing of the barrier with $a=b=1$,
initial energy $K=0.41$ and extraction potential $V_0=100$ Volts.} 
\label{fig:fig4} \end{figure} 

\begin{figure} \protect\caption{Evolution of the kinetic energy 
during the crossing of the barrier with $a=1$, 
$b=10$, initial energy $K=0.91$ and extraction potential 
$V_0=1000$ Volts.} \label{fig:fig5} \end{figure} 

\begin{figure} \protect\caption{Classical probability of tunneling
$P(b)$ and quantum tunneling for four different extraction 
potentials $V_0$.} \label{fig:fig6} \end{figure}

 
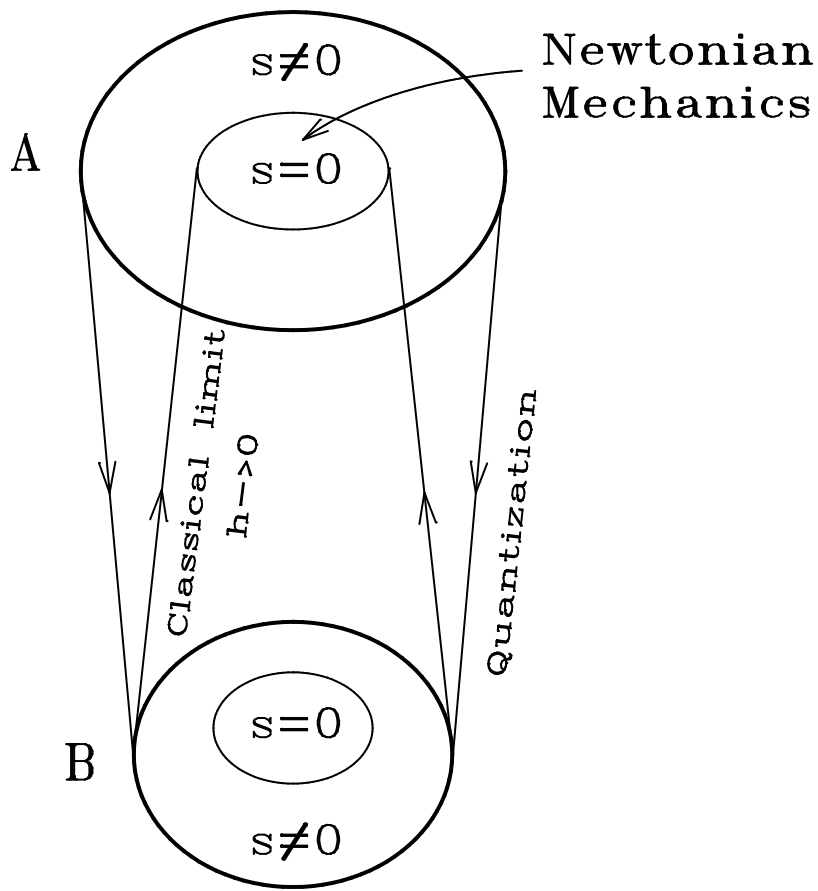
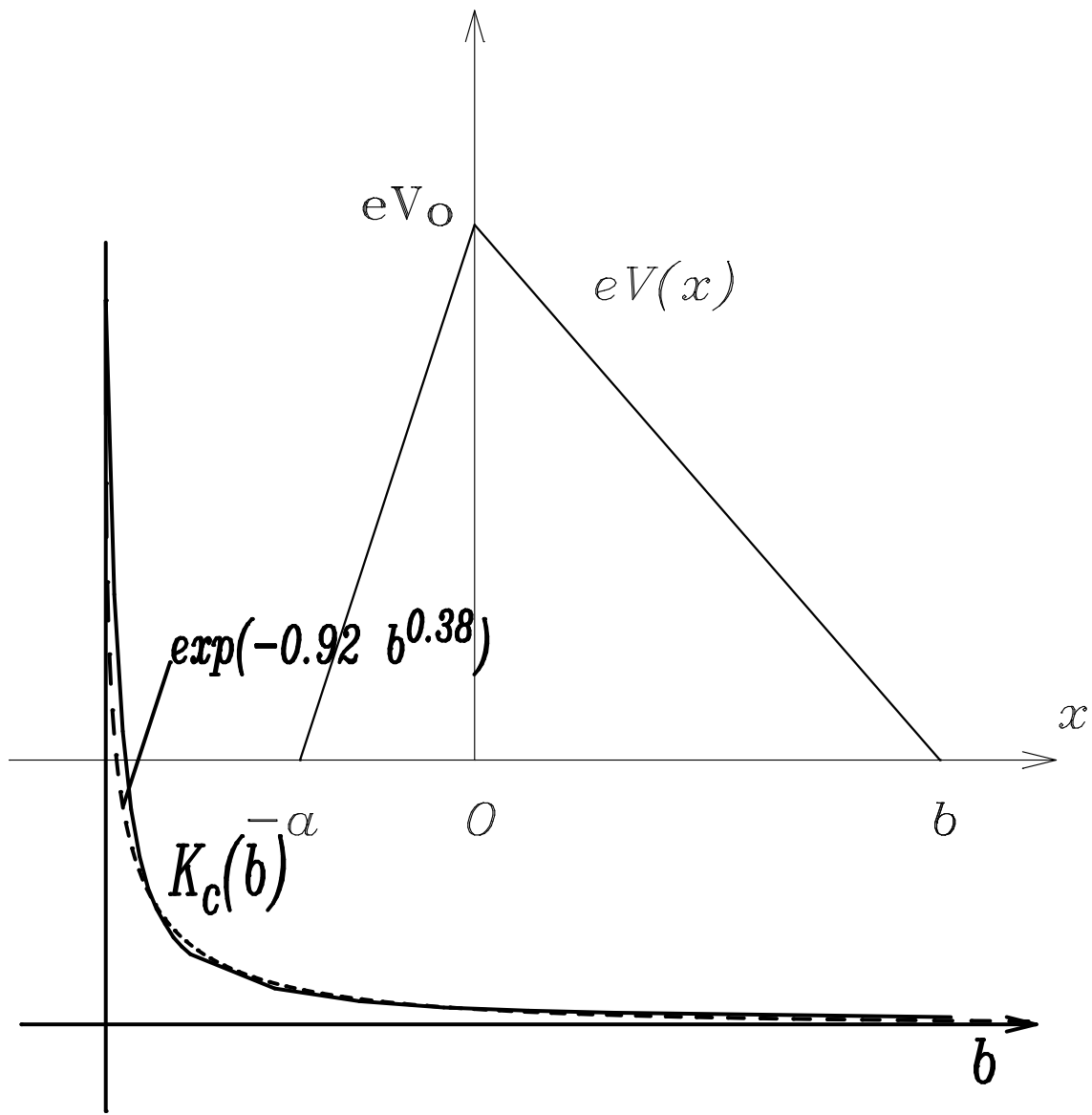
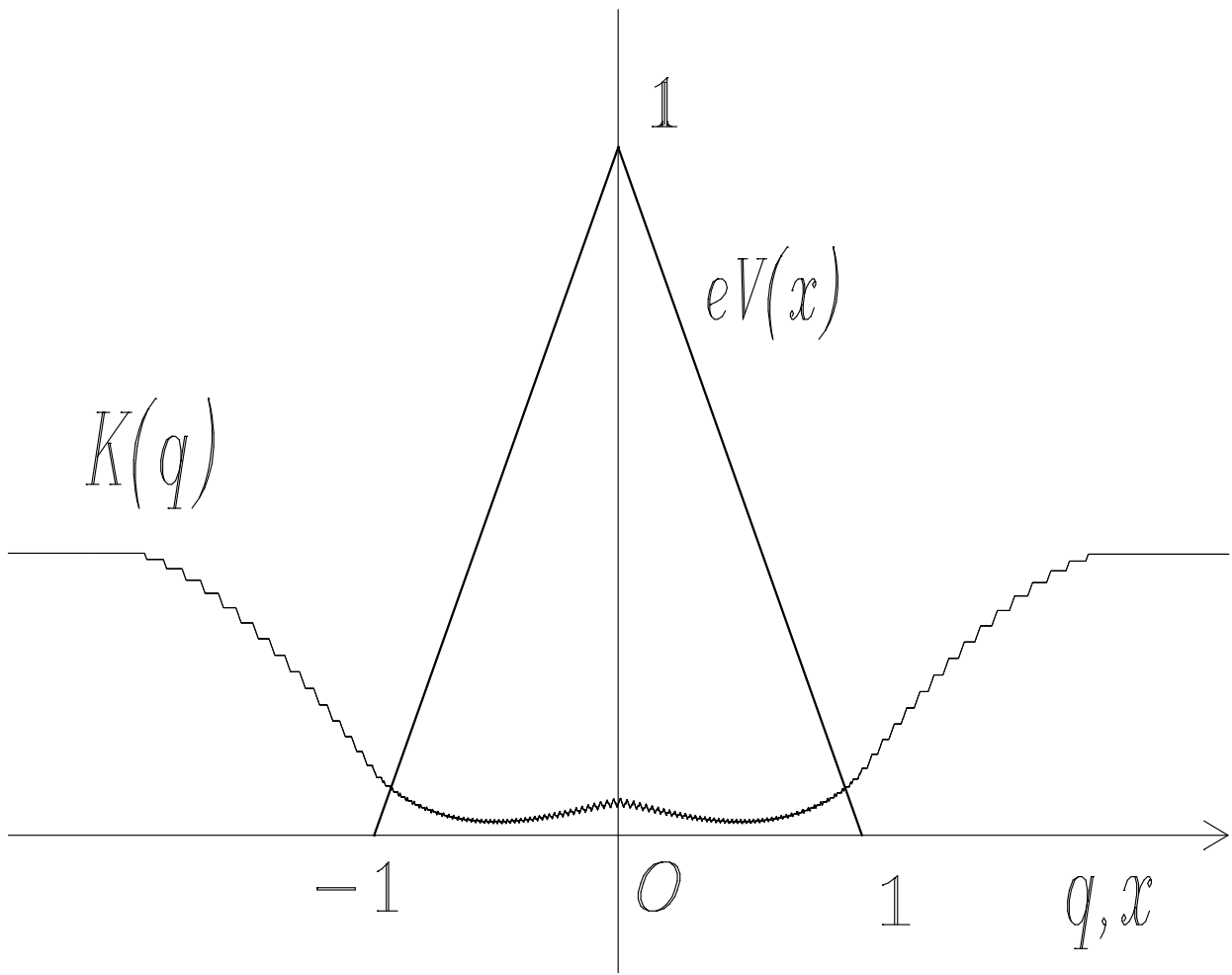
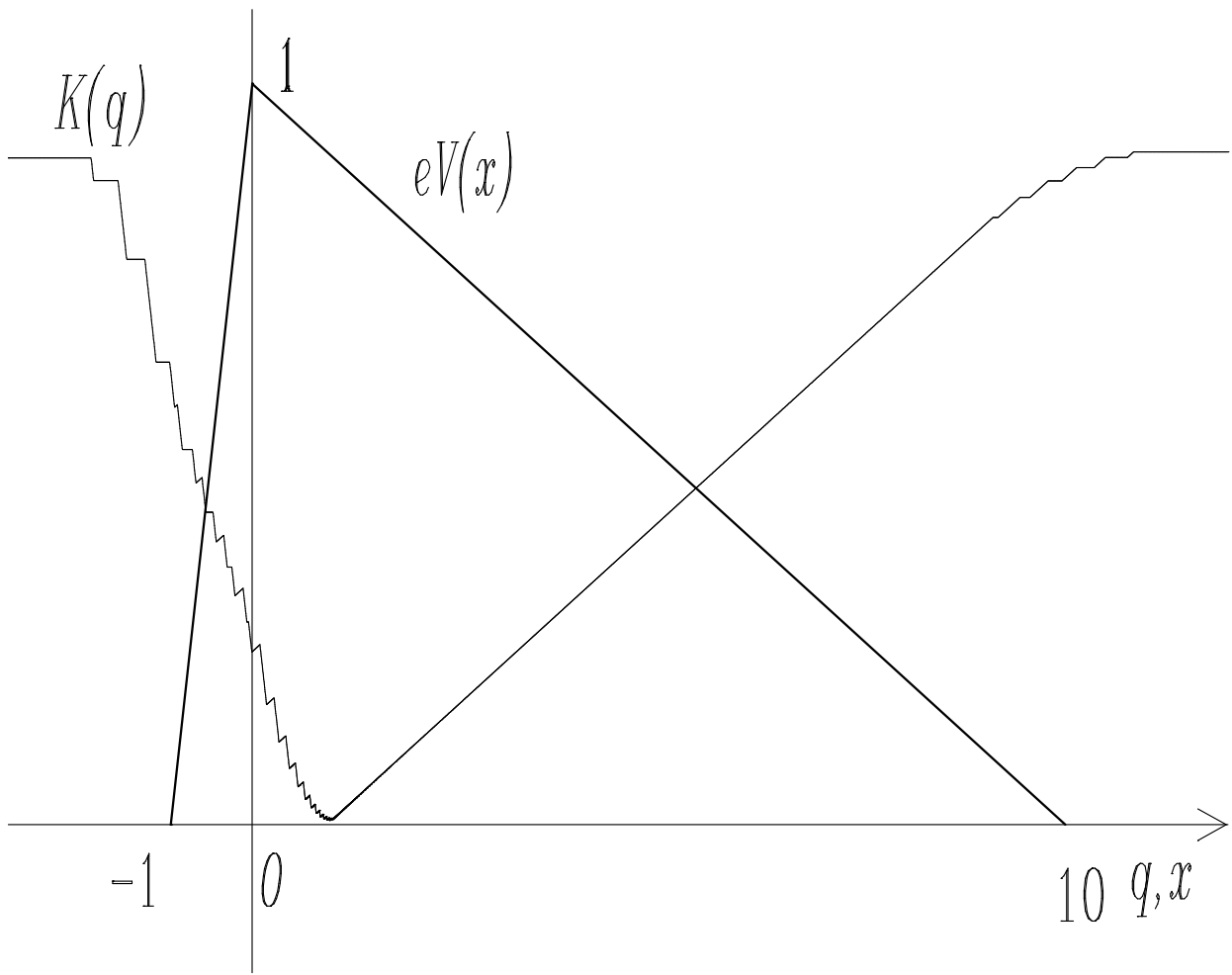
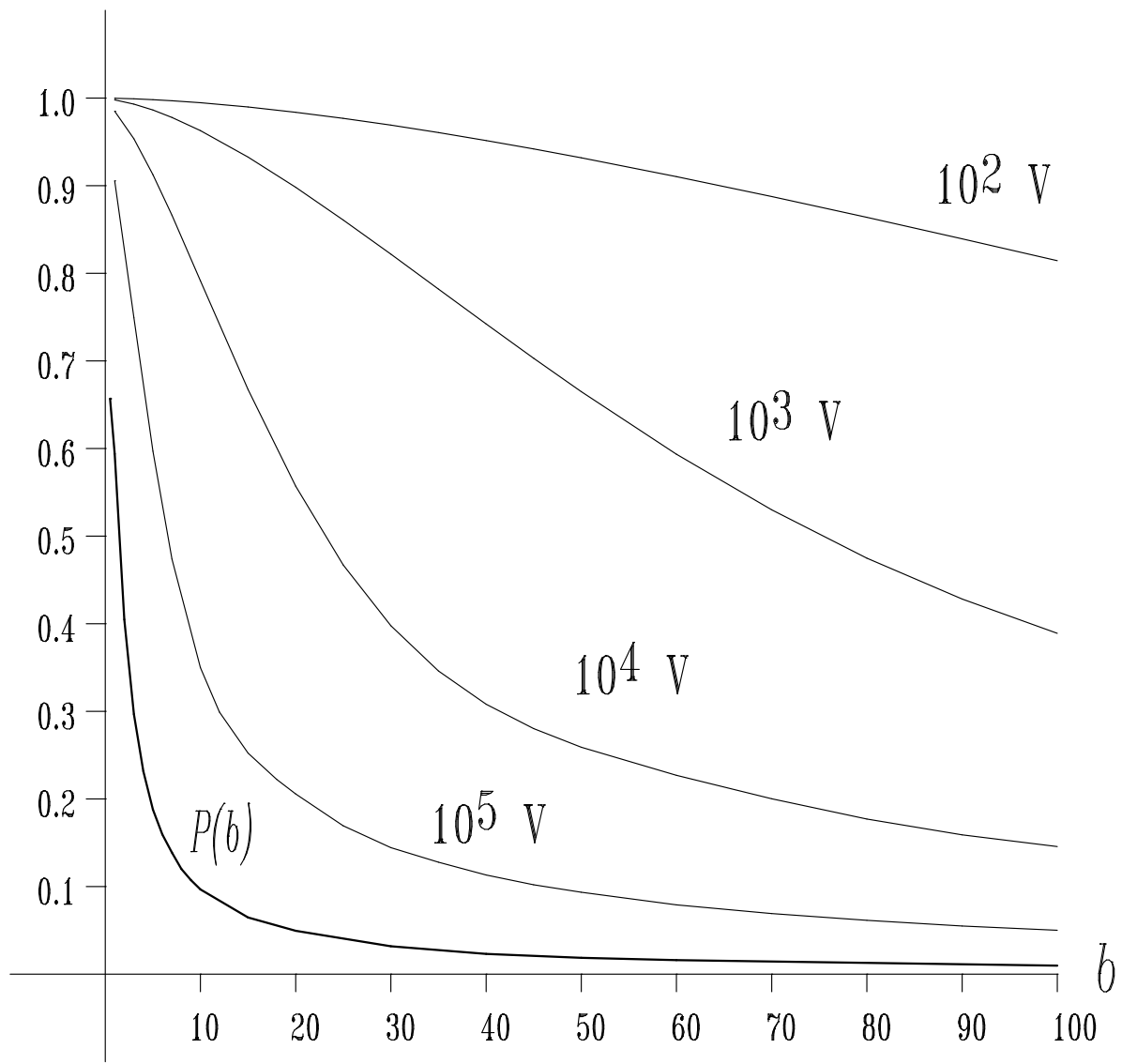
\end{document}